\documentclass{article}
\usepackage{graphicx} 

\title{Consideration for a gamma source at EupraXia       }
\author{Luca Serafini $^{1}$, Vittoria Petrillo $^{{2,1}}$*, Alessio DelDotto, $^{3}$ \\ Illya Drebot $^{{1}}$, Anna Giribono $^{3}$, Andrea Ghigo $^{3}$,  Stefano Romeo $^{3}$, \\  Andrea Renato Rossi $^{1}$,  Cristina Vaccarezza $^{3}$, Fabio Villa $^{3}$ \\and Massimo Ferrario $^{3}$}
\date{September 2025}

\begin{document}

\maketitle


$^{1}$ \quad INFN-Section of Milan, Via Celoria,16 20133 Milano (Italy)\\
$^{2}$ \quad Università degli Studi di Milano, Via Celoria,16 20133 Milano (Italy)\\
$^{3}$ \quad INFN-LNF, Via E. Fermi, 54, Frascati,00044, Italy

Correspondence: Vittoria.Petrillo@mi.infn.it


\abstract{In this paper, we present the potentiality and the application of a Compton source driven by the electron beam of EuPRAXIA, in interaction with a commercially available infrared laser.}

Thomson scattering; Compton scattering; Synchrotron radiation, X-rays; Radiation sources.

\section{State of the art and application of gamma-rays}

The interest in the production of monochromatic $\gamma$-rays first came from studies on Nuclear Resonance Fluorescence (NRF) \cite{Kneiss}. NRF is a technique consisting of the excitation of nuclear states by absorption of resonant radiation, with subsequent decay of these levels by re-emission of equivalent real photons. The NRF method with 100 \% linearly polarized photon beams allows for the measure of spins, parities, branching ratio and decay widths of the excited states. It represents an important tool for the study of low-multipolarity ground-state transitions with large partial widths and for measuring the dipole transitions \cite{Howell}. 
Among all the X/$\gamma$ Compton sources operating worldwide \cite{stateofart}, the actual gamma source of reference for these studies is HI$\gamma$S at the Duke University \cite{Higs}. A program of nuclear photonic experiments is also in progress at the Compton source at NewSUBARU, Osaka \cite{NewSUBARU}.
Nuclear resonance experiments have been carried out at HI$\gamma$S since 2001, first with gamma rays with energy from 5.4 to 6.5 MeV \cite{Pietralla1,Fransen}. 
An upgrade of the energy permitted them to investigate the so-called Pygmy Dipole Resonances (PDR) \cite{Savran} and the core polarization dipole modes, which have been observed as a clustering of states close to the neutron threshold at excitation energies $E_x \simeq$ 5.5-8.6 MeV \cite{Pietralla2,Tonchev2,scheck, Tonchev4}. 
Other experiments at HI$\gamma$S studied the mechanism of excitation of isomeric $h_{11/2}$ states in nuclei around the closed shell at N = 82 \cite{Angell,tonchev3}. 
The dipole strength of the nuclide $ ^{66} Zn$ was also studied in photon scattering experiments using quasi mono-energetic and linearly polarized photon beams of 30 energies within the range of 4.3 to 9.9 MeV \cite{Fransen2}.

The Giant Dipole Resonance at 8-16 MeV in the target nuclei $^{138}Ba$, $^{140}Ce$, $^{142}Nd$ and $^{241}{Am}$ has been excited, and, following neutron emission, the $\gamma$-ray cascades leading to the isomeric state and the ground state have been observed \cite{Pietralla2,Tonchev6}. 
Understanding the electroweak response of $ ^4He $ nuclei in the giant dipole region is essential to resolve the neutrino nucleosynthesis process in supernovae. At NewSUBARU, two photodisintegration reactions, $ ^4He $  ($\gamma$,n) $^3He$ and $ ^4He $ ($\gamma$,p)$ ^3H$, in the energy range of $E_\gamma$ = $23.0–33.3$ MeV have been simultaneously measured \cite{elio, elio2}.
Facilities with a quasi-monoenergetic $\gamma$ -ray beam are also ideal tools for studying photodisintegration reactions which are relevant for the nucleosynthesis of heavy and relatively proton-rich nuclei, the so-called p nuclei, in the $\gamma$-process. 
The origin of the nuclei lying on the proton-rich side of the valley of  $\beta$-stability between $^{74}Se$ and $^{196}Hg$ can be explained by burning of the pre-existing s- and r-nuclei in stellar environments of high enough temperature. At these high temperatures, high-energy photons initiate a sequence of reactions- namely ($\gamma$, n), ($\gamma$, p), and ($\gamma,\alpha$ )- that leads to the production of the p nuclei. Exploiting a further increase of the energy of quasi-monochromatic $\gamma$-beams, the cross section for the production of the $^{180m} Ta$ isomer was determined in the energy region from 8 to 15 MeV \cite{Tonchev}, improving the determination of the corresponding stellar photodisintegration rate of $^{181}Ta$, which directly influences the p-process production of $^{180}Ta$. 
Another nuclear application, important for national security issues, is the detection of a high concentration of nitrogen and oxygen with a low concentration of carbon, indicating the presence of an explosive material. A mono-energetic ($\Delta$E = 130 keV) and polarized $\gamma$-beam  was used to resonantly excite the $^{14}N$ to the level at 9172.2 keV. Then, the NRF technique was used to detect the $\gamma$-rays resonantly scattered from this level. 
Going towards higher energies, the production of 120 MeV gamma-ray beams was first demonstrated in 2021 at HI$\gamma$S \cite{120MeV}.
120 MeV gamma-ray beams for users were first produced for an experimental test run using a $^{12}C$ target, as a part of the electromagnetic polarizability research program. 
The next interesting higher-energy range is from 130 MeV to 150 MeV, which will create new opportunities for nucleon spin-polarizability measurements.

In this paper, the performances of a Compton Gamma-ray source based on the EuPRAXIA plasma-driven beam \cite{TDR} are analyzed in the photon energy range interesting for nuclear photonic applications and compared with the state of the art.
The wide tuning of such source and the possibility of fundamental experiments is underlined in connection with the extension of the EuPRAXIA electron beam energy in future upgrades.

\section{A Gamma-ray  source at EuPRAXIA}

EuPRAXIA \cite{sito} is one of the projects of the European Strategy Forum on Research Infrastructures (ESFRI) Roadmap of 2021. It is the first European project developing a dedicated particle accelerator research infrastructure based on novel plasma acceleration concepts driven by innovative technologies. The Italian pillar of EuPRAXIA \cite{TDR} will be host in Frascati at the INFN-LNF and will explore the potentiality of a plasma beam-driven accelerator. A combed electron beam, constituted by a driver and a witness one, will rely on the most compact RF technology available, the X-band structures. The electron beam accelerated in the linac up to the energy of 1 GeV at 100 Hz will then be injected into a plasma acceleration stage. The nominal working point of the first phase envisages the exit of the electrons from the linac at 500 MeV and the gain of a further amount of about 500 MeV of energy in the plasma cell. First phase user applications for EuPRAXIA$@$SPARC$\_$LAB will focus on a $3-10$ nm free-electron laser, while further upgrade will foresee the extension to 400 Hz of repetition rate, the implementation of an inverse Compton scattering photon source, high-energy positron and test beams.  In future upgraded stages, it is foreseen that the energy will be further increased, approaching the ideal target of 5 GeV.

The X/$\gamma$-rays inverse Compton sources exploit the scattering between electrons and radiation. 
An electron beam, in interaction with a counter-propagating laser pulse, produces X/gamma rays with energy \cite{Theory, Ranjan}:
\begin{equation}
E_{X/\gamma}=E_{L} \frac{\gamma^2 (1-\beta cos\alpha)}{\gamma^2 (1-\beta cos\theta)+\frac{X}{4}(1-cos(\alpha+\theta))},
\end{equation}
where $E_{X/\gamma}$ is the radiation photon energy, $E_L$ is the energy of the laser photons, and $X=\frac{4E_eE_{L}}{(m_0c^2)^2}$ is the recoil factor.
Moreover, $\alpha$ is the laser injection angle, which should be close to $\pi$ (head-on collision) and $\theta$ is the photon emission angle.
For head-on injection, the photon energy on axis turns out to be:
\begin{equation}
    E_{X/\gamma}= \frac{4 \gamma^2 E_L}{1+X}.
\end{equation}

As already said, EuPRAXIA \cite{TDR}, in its first stage, will produce electron beams with  energies up to 1 GeV, in the configuration of plasma beam-driven acceleration.  The tunability of a Compton source for a few energies of the EuPRAXIA electron beam in the various stages is shown in Table \ref{tab:tune}. With the nominal EuPRAXIA beam at energy of about 1 GeV, the source generates photons at energy of 18 MeV, just in the center of the interval between 10 and 25 MeV, typical of the Giant Dipole Resonance. Changing the electron beam energy by 20\% around 1 GeV, namely from 0.8 to 1.2 GeV, well inside the operating tolerance margins of all elements in the electron beam line, the photon energy can indeed be tuned from 11.5 to 26.7 MeV, covering most of the Giant Dipole Resonance energy range. We can also observe that operating with lower accelerating gradient, or extracting the electron beam before the end of the line, enables to make available lower electron energies, with the possibility of exploring the region of photon energy $<$ 10 MeV. Since the EuPRAXIA linac is projected to operate up to 1 GeV, and the plasma accelerating cell doubles the energy, the facility, in its first phase, can, in principle, push the electron energy up to 2 GeV. The photon energy, in turn, will reach 74 MeV. The elongation of the plasma acceleration cell by a few centimeters may increase further this value up to 90 MeV, without substantial changes in the facility layout. The upgrade towards 5 GeV, foreseen (see \cite{TDR}, last chapter) in the future, will permit to push the radiation photon energy to 450 MeV, passing through all intermediate values.

The recoil factor remains always below 1, so a classical approach is appropriate. Only with the most energetic electron beam and using the harmonics of the laser, quantum effects could be highlighted \cite{Vittoria}. 
The number of emitted photons per shot can be roughly evaluated with the following simple formula, based on the source luminosity:
\begin{equation}
    N_{X,\gamma}= \sigma_{Th}\frac{ N_e N_L}{2\pi \sigma_x^2}
    \label{eq:lum}
\end{equation}
where $\sigma_{Th}$=6.65 $10^{-29} m^2$ is the Thomson cross section,
\begin{table}
    \centering
 
    \begin{tabular}{|c|c|c|c|c|}  
    \hline
    
           &Electrons& Laser& Recoil&Radiation \\ 
           &GeV& eV& &MeV\\
           \hline
           $1^{st}$ phase& 1 &1.21 &0.018&18\\ 
           
           $1^{st}$ phase&0.5-0.8&1.21&0.005-0.0092&5-11.5\\
          
$1^{st}$ phase& 0.8-1.2 & 1.21&0.015-0.02&11.5-26.7\\
Upgrade 1& 1.2-2.2 &1.21 &0.02-0.04&26.7-87\\
 Upgrade 2&2.2-5 & 1.21&0.04-0.1&87-450\\
 Upgrade 2&2.2-5 & 2.42&0.08-0.2&175-900\\
 
 \hline

    \end{tabular}
    
    \caption{Compton source tunability}
    \label{tab:tune}
\end{table}

The typical characteristics of electrons under such conditions are listed in
Table \ref{tab:electrons}. In particular, two energy values of interest have been reported and start-to-end simulated. The first one is 1 GeV, the working point widely described in the EuPRAXIA TDR (see \cite{TDR}, Chapter Beam Dynamics), obtained with the injection of a 500 MeV beam from the linac to the plasma acceleration stage. The second working-point we have studied has electron energy at 2.2 GeV and belongs to a further upgraded phase.
For an eventual Compton source, a high-charge electron beam is the most suitable one. Here, we considered 50 pC of charge, 100 Hz of repetition rate, 0.55 mm mrad of normalized emittance $\epsilon_{x,y}$ in both planes, $4 10^{-3}$ of relative energy spread $\Delta E/E$, according to the values obtained by the start-to-end simulations presented in the EuPRAXIA TDR for operations already planned in the first phase of the project \cite{TDR}. In future upgrades, electron beams with 100 pC at $E_e$=2.2 GeV and 400 Hz are supposed to be produced at EuPRAXIA, with similar values of the other parameters.

\begin{table}
    \centering
 
    \begin{tabular}{|c|c|c|c|}  
    \hline
    &&$1^{st}$ phase&Upgrade\\\hline
           Energy& GeV& 1&2.2 \\
  $N_e$& $\times 10^8$&3.125& 6.25\\  
 Charge& pC& 50& 100\\
 $\sigma_z$& mm& 0.006& 0.006\\  
$\beta_x=\beta_y$& m&0.6& 0.1\\
 $\epsilon_x$=$\epsilon_y$& mm mrad&0.55& 0.55\\
 $\sigma_{x,y}$&$\mu m$& 10& 10\\
 $\sigma_E$/E& &0.004&0.002\\
Rep rate& Hz& 100 &400\\
\hline

    \end{tabular}
    
    \caption{Electron beam parameters}
    \label{tab:electrons}
\end{table}

The main characteristics of the laser pulse are summarized in Table \ref{tab:laser}.
\begin{table}
    \centering
 
    \begin{tabular}{|c|c|c|c|}  
    \hline
           &&$1^{st}$ phase&Upgrade\\\hline
           Energy& J& 0.5&0.5 \\  
 Wavelength& nm& 1030& 1030\\
 $\sigma_z$& mm& 0.5&0.5\\  
Rayleigh length& mm&1.4&1.4 \\
 angle& rad&0&0 \\
 $\sigma_{x,y}$& $\mu m$ &10&10\\
Rep rate& Hz& 100 &400\\\hline

    \end{tabular}
    
    \caption{Laser  parameters}
    \label{tab:laser}
\end{table}
\begin{table}
    \centering
 
    \begin{tabular}{|c|c|c|c|}  
    \hline
    &&$1^{st}$ phase&Upgrade\\\hline
Electron energy& GeV& 1&2.2 \\ 
Gamma edge&MeV&18&87\\
Minimum bandwidth&$\%$&1&1.25\\
X photon/shot & $\times10^7$ &3.9 &7.2\\
X photon/s & $\times10^9$&3.9 &28.8\\
Collimation angle& $\mu rad$ &150&70 \\
Bandwidth& $\%$ & 3& 3\\
Collimated phot./shot&$\times$ $10^6$& 4 &8\\
Collimated phot./ s&$\times$$10^8$/s& 4 &32\\
Spectral density& ph/s/1eV&300&850\\

\hline
    \end{tabular}
    
    \caption{X-rays parameters}
    \label{tab:X-ray}
\end{table}
\begin{figure}
    \centering
    \includegraphics[width=0.8\linewidth]{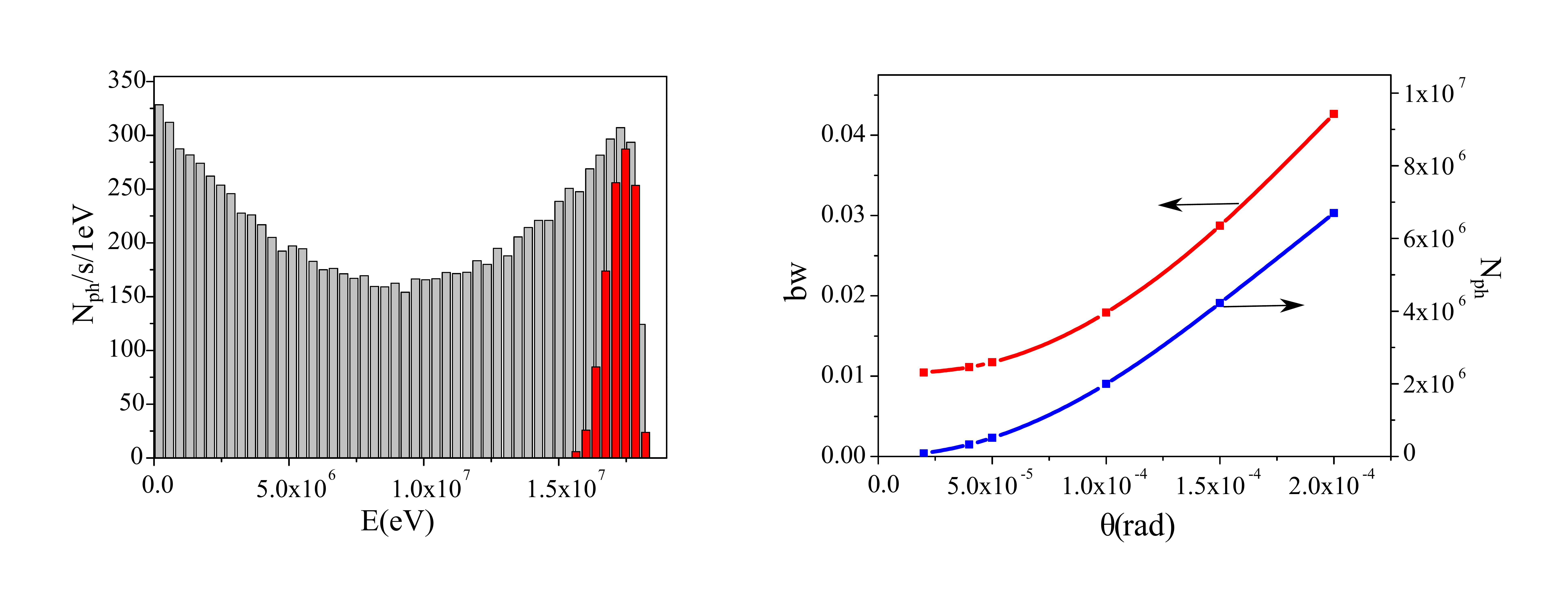}
    \caption{Total (in grey) and collimated (in red) spectrum of gamma-ray for electron energy at 1 GeV. Photons collected in 150 $\mu$rad, for a bandwidth of 3\%. Bandwidth and number of photons as function of collimation angle.}
    \label{fig:gam1}
\end{figure}
\begin{figure}
    \centering
    \includegraphics[width=0.8\linewidth]{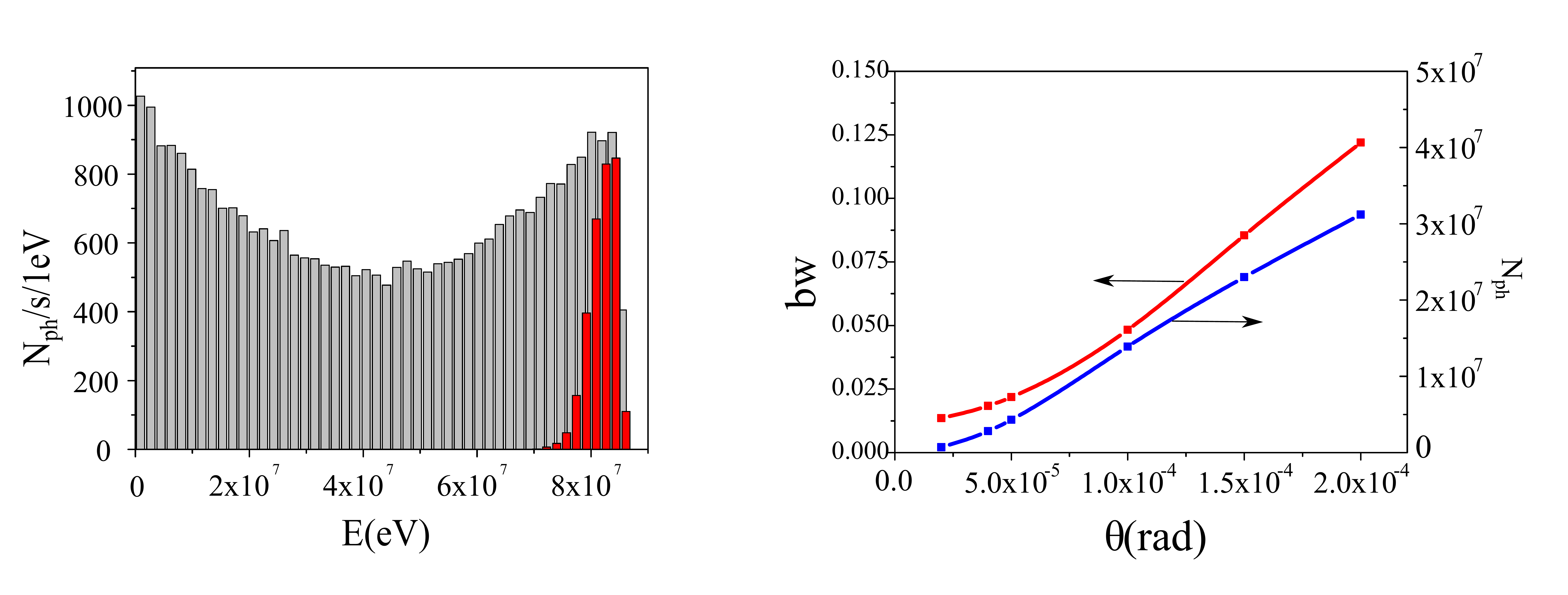}
    \caption{Total (in grey) and collimated (in red) spectrum of gamma-ray for electron energy at 2.2 GeV. Photons collected in 70 $\mu$rad, for a bandwidth of 3\%. Bandwidth and number of photons as function of collimation angle.}
    \label{fig:gam2}
\end{figure}

As collision laser we have thought of a Yb:Yag 100 Hz high quality laser system, synchronized to an external photo-cathode laser and to the RF system to better than 1 ps time jitter. This technology is state-of-art and currently used in other Compton sources as, for instance, in the project STAR \cite{STAR}. Both linear and circular polarization at almost 100\% should be available. 
400 Hz of repetition rate is expected in future upgrades. The laser characteristics are listed in Table \ref{tab:laser}.
An evaluation of the $\gamma$-ray outcome based on formula \ref{eq:lum} for Q=50 pC, 0.5 J of laser energy, 13 $\mu m$ of transverse e-beam rms spot dimension and 100 Hz of repetition rate gives a number of photons per shot of about $ 5 \times 10^7$ , corresponding to  $5 \times 10^9$ photon per second. In the upgraded case, by using 100 pC, 0.5 J of laser energy, same dimension as before and 400 Hz of repetition rate, we obtain $9.8\times 10^7$, which means $3.9 \times 10^{10}$ photon per second.
More precise estimations of the radiation outcome have been made by using a simulation code well established in the Compton community, namely the code CAIN \cite{CAIN}, which also takes into account the propagation of both beams and the possible losses in photon flux due to transverse inhomogeneities and diffraction effects.
The $\gamma$-radiation obtained in this way is presented in Table \ref{tab:X-ray} for both cases, first phase and upgrade. Figures \ref{fig:gam1} and \ref{fig:gam2} give at left the total and collimated spectra at 3\% of bandwidth and the collimated radiation bandwidth and photon number as a function of the acceptance angle. A total photon number of 3.9 $10^7$ and 7.2 $10^7$ photons per shot is achieved respectively in the two cases. The low values of emittance and energy spread obtained in the EuPRAXIA gun and linac, and not deteriorated in the last module due to the method of beam-driven plasma acceleration, guarantee a low minimum bandwidth threshold at zero acceptance of 1\% and 1.3\% respectively, in agreement with the residual value evaluated by Equation 14 of Ref. \cite{Theory} . 

The polarization of the radiation reflects the polarization state of the laser \cite{pol}. 
A collimator, selecting a bandwidth of about 3 $\%$, is placed on the $\gamma$-rays trajectory, for a collimated photon number of one order of magnitude less. 
The spectral density of the total and collimated radiation is presented in Figure \ref{fig:gam1} for 18 MeV and in Figure \ref{fig:gam2} for 87 MeV of Compton edge.
The peak spectral densities reach 300 and 850 photons per second and per 1 eV of bandwidth, respectively.
Both cases, the one with the lowest photon energy of 18 MeV and the other with the largest energy of 87 MeV, are competitive with respect to the reference source HI$\gamma$S.

\section{Scientific case of the EuPRAXIA gamma source }

The tunability of the present gamma source could enable for a large number of experiments. The expected values, few of them shown in Table \ref{tab:X-ray}, suit the requirements of a wide nuclear photonics research program \cite{Howell}.
In particular, energies between 5 and 10 MeV, with 3\% of bandwidth and $10^9$ $\gamma$/s with 95\% of linear/circular polarization, are required in photofission experiments for the excitation of bound and unbound individual nuclear states. The operation of the source with the linac-only acceleration could be foreseen for such low photon energy ($<11 $MeV) /high-flux ($>$ 3 $10^8$ $\gamma$/s) working points.
The range between 10 and 25 MeV, matching exactly the nominal primary regime of operation of EuPRAXIA, coincides almost precisely with the Giant Dipole Resonance range. The bandwidth and flux values also fit the requirements of such experiments.  
In the photon energy range from 60 to 120 MeV, the source can be used for experiments on neutron scalar polarizabilities from $^4He$. Fluxes on target from $10^8$ to $10^9$ $\gamma$/s, with relative energy resolution of 3\% FWHM, are required.
From 100 to 140 MeV, achievable with electron energy up to 2.75 GeV in a further phase of upgrade, neutron scalar polarizabilities from deuteron and neutron spin polarizabilities from polarized $^3He$  can be tested with flux on target of 5 $\times$$10^8 \gamma$/s. The large degree of polarization of EuPRAXIA, estimated in more than 95\%, together with the relative energy resolution of 2\% FWHM, meets the constraints of such measurements.
Again in this further upgrade, during the endeavors to achieve 5 GeV of electron energy, photon energies from 150 to 300 MeV can be produced and used in the field of experiments on proton spin polarizability.
Flux on target of $10^9$ $\gamma$/s, with relative energy resolution of 2\% FWHM
with linear and circular polarization  at $>$95\% are required. 
Finally, the largest electron energy, combined with the laser harmonics, pushes the system toward the quantum regime, where fundamental studies can be proposed \cite{Vittoria}.

\section{Conclusion}
We have shown that even in the first phase of the EuPRAXIA project, the plasma accelerated electron beam, in interaction with a commercially available infrared laser, is able to drive a competitive Compton source.
The basic photon energy range is suitable for nuclear photonics applications, while flux and spectral density are comparable with those of the state of the art source HI$\gamma$S in the same spectral range, enabling a wide range of nuclear photonics experiments.
Larger electron energy combined with the laser harmonics pushes the system toward the quantum regime, where fundamental studies can be proposed \cite{Vittoria}.
Finally, by extracting the electron beam with lower energy (150-200 MeV) photons of 200-300 keV of energy could be generated, an energy range suitable for FICS studies \cite{FICS}.

\end{document}